\documentstyle[prd,aps,preprint,epsfig]{revtex}
\tighten

\begin{document}
\draft
 
\pagestyle{empty}

\preprint{
\noindent
%\begin{minipage}[t]{3in}
%\begin{flushleft}
%\today \\
%\end{flushleft}
%\end{minipage}
\hfill
\begin{minipage}[t]{3in}
\begin{flushright}
%LBL--xxxxx \\
%UCB--PTH--01/xx \\
%hep-ph/01xxxxx \\
February 2001
\end{flushright}
\end{minipage}
}

\title{EXTRAPOLATION OF $K\rightarrow\pi\pi$ DECAY AMPLITUDE}

\author{
Mahiko Suzuki
%\thanks{Work supported in part by the Director, Office of Energy
%Research, Office of High Energy and Nuclear Physics, Division of High
%Energy Physics of the U.S. Department of Energy under Contract
%DE--AC03--76SF00098 and in part by the National Science Foundation under
%grant PHY--95--14797.}
}
\address{
Department of Physics and Lawrence Berkeley National Laboratory\\
University of California, Berkeley, California 94720
}

%\thanks{Work supported by the Department of Energy under Contract
%DE--AC03--76SF00515.}

%\date{\today}
\maketitle

\begin{abstract}

We examine the uncertainties involved in the off-mass-shell extrapolation 
of the $K\rightarrow \pi\pi$ decay amplitude with emphasis on those aspects 
that have so far been overlooked or ignored. Among them are initial-state 
interactions, choice of the extrapolated kaon field, and the relation 
between the asymptotic behavior and the zeros of the decay amplitude. 
In the inelastic region the phase of the decay amplitude cannot be 
determined by strong interaction alone and even its asymptotic value
cannot be deduced from experiment. More a fundamental issue is intrinsic 
nonuniqueness of off-shell values of hadronic matrix elements in general.  
Though we are hampered with complexity of intermediate-energy meson 
interactions, we attempt to obtain a quantitative idea of the uncertainties 
due to the inelastic region and find that they can be much larger than 
more optimistic views portray. If large uncertainties 
exist, they have unfortunate implications in numerical accuracy of 
computation of the direct CP violation parameter $\epsilon'$.

\end{abstract}
%\pacs{}
\pacs{PACS number(s): 13.20.Eb, 11.55.Fv, 13.75.Lb, 11.10.Jj}
%\newpage
\pagestyle{plain}
\narrowtext

\setcounter{footnote}{0}

\section{Introduction}

Accurate computation of the long-distance QCD corrections for the
decay $K\rightarrow\pi\pi$ is important in testing the underlying 
mechanism of CP violation with the $\epsilon'$ parameter. 
When one obtains a real amplitude for $K\rightarrow\pi\pi$ by any
method of calculation, its validity is limited to the unphysical 
region below $\pi\pi$ rescattering. On the kaon mass shell one
must supply a phase by taking account of 
rescattering\cite{Truong,Isgur}. However, rescattering affects 
a magnitude of amplitude in general. While the phase of amplitude 
arises solely from elastic rescattering in the case of $K\rightarrow 
\pi\pi$, the magnitude of amplitude is affected by inelastic 
scattering as well. Many attempts have been made to incorporate
the final-state interaction (FSI) in the $K\rightarrow\pi\pi$
decay amplitudes, most recently by Palante and Pich\cite{Pich}
among others.
    
The final-state interaction (FSI) theory was first formulated in 
potential scattering in the 1950's\cite{Fermi}, then extended 
with dispersion relation to relativistic particle physics, often 
using the Omn\`{e}s-Muskhelishvili (OM) representation\cite{OM}.
When our interest was in very low energy phenomena, it was a good
enough approximation to discard all but elastic FSI rescattering. 
In particle physics, however, inelastic rescattering can be 
potentially important to the magnitude of amplitude even when it does 
not contribute to the phase. If we knew the values of the phase
at all energies, we could compute the magnitude with the OM representation
apart from factors of zeros.
However, the phase of a decay amplitude above the inelastic 
rescattering threshold has nothing to do with the inelastic scattering 
phase shift of strong interaction. It depends on decay operators even when
all quantum numbers are the same. 

In addition to inelastic rescattering, strong interactions take 
place in the initial state before $K$ decays. Though it does not 
contribute to the phase, the initial-state interaction (ISI) generates 
$p_K^2$ dependence for the off-shell amplitude just as the FSI does.
The existing calculations of the FSI largely neglect the ISI effect. 
Once the inelastic FSI and the ISI are included, the OM representation is
by no means 'universal'' contrary to the statement often found in recent
literature.

Extrapolation off mass shell involves a few more basic issues that have 
not been discussed. One is the asymptotic behavior of a 
decay amplitude far off mass shell, which is directly related to the number 
of zeros and the asymptotic phase $\delta(\infty)$ of the amplitude. 
In the elastic approximation, we may follow potential scattering theory 
and determine the asymptotic behavior by Levinson's theorem\cite{Levinson}. 
In relativistic physics which contains inelasticity, Levinson's 
theorem does not hold. Therefore we must determine in 
one way or another the value of $\delta(\infty)$ 
and the number of zeros and incorporate them in the OM representation. 

Another issue is nonuniqueness of off-shell amplitudes, 
which is inherent in quantum field theory. According to a general  
theorem\cite{Kame}, on-shell amplitudes are unique no matter what operator 
one may choose for a particle field as long as it has a correct wave 
function renormalization and a right set of quantum numbers on mass shell. 
However, off-shell amplitudes can depend on choice of particle fields,
which is by no means unique in the case of hadrons. 
In particular, the asymptotic behavior is sensitive to the choice.
In order to write the OM representation uniquely, therefore, we must 
first define the kaon field and then determine the asymptotic behavior 
and the number of zeros of the decay amplitude.
 
Since we think that many of these basic issues in the off-shell 
extrapolation have been either improperly treated or entirely 
ignored in literature, we attempt to expose them and obtain 
some quantitative idea of the uncertainties associated with them 
in this paper. While a recent criticism\cite{Buras} concerns mostly technical
aspects of Ref.\cite{Pich}, we focus on more basic aspects of the FSI
theory\cite{T2}.

The paper is organized as follows:
In Sec. II, we clarify distinction between the FSI and the ISI.
For this purpose we write the $K\rightarrow\pi\pi$ amplitude first by 
the dispersion relation for scattering of a weak spurion off an on-shell 
kaon. In this dispersion relation, the FSI generates the $s$-channel
singularities while the ISI contributes to the $u$-channel singularities. 
Then we write the OM representation by dispersing the kaon mass $p_K^2$ and 
show that both the FSI and  the ISI contribute to the right-hand cuts 
in the OM representation. In Sec. III, we relate the number of 
zeros of amplitude to the asymptotic phase and magnitude in the OM 
representation in general. In Sec. IV, after reviewing the inherent
ambiguity of an extrapolated particle field in field theory, we study 
the asymptotic behaviors of the $K\rightarrow\pi\pi$ amplitudes in QCD by 
defining the off-shell kaon field by the partially-conserved axial-vector 
current (PCAC) relation. 
In Sec. V, we interpolate the phases for the dominant $\Delta I
=\frac{1}{2}$ amplitude and for the $\Delta I = \frac{3}{2}$ 
amplitude between the inelastic threshold and the high energy limit.
Then we examine in detail the case that the decay amplitudes have the
smallest number of zeros including the well-known low-energy zero.
In this particular case we make  
quantitative discussion on the extrapolation effect by giving 
illustrative estimates of how much variation occurs in the extrapolation  
from $p_K^2=m_{\pi}^2$ to $m_K^2$. We find that the ISI and the inelastic 
FSI are not negligible even in the case of a single zero.  In more general
cases, uncertainty is larger. The source of the largest uncertainty is 
in the number of zeros and their locations.

\section{Dispersion relations}

The decay matrix element for $K\rightarrow\pi\pi$ is written as
\begin{equation}
     M = \langle \pi(p_a)\pi(p_b)^{out}|H_W(0)|K(p_K)\rangle.
\end{equation}
In actual computation an effective decay operator $c_i(\mu){\cal O}_i$
is chosen for $H_W$, which incorporates short-distance QCD effects in 
$c_i(\mu)$ by renormalization group. Since the interaction above the 
energy scale $\mu$ is included in $c_i(\mu){\cal O}_i$, we ought to truncate 
dispersion integrals accordingly, as we shall discuss later. 

\subsection{Spurion scattering}

    Treat $H_W$ as the source of spurion $S$ by setting $\Box S = H_W$ and
write a dispersion relation for the on-shell scattering,
\begin{equation} 
      S(p_s) + K(p_K) \rightarrow \pi(p_a) + \pi(p_b), \;\;\;(p_s^2=0)
              \label{process}
\end{equation}
with the Mandelstam variables,
          $s=(p_s+p_K)^2=(p_a+p_b)^2$, 
          $t=(p_K-p_a)^2=(p_s-p_b)^2$,
and $u = (p_K-p_b)^2 = (p_s-p_a)^2 = m_K^2+2m_{\pi}^2-s-t$. The decay 
amplitude $M(s,t)$ at $t<(m_K+m_{\pi})^2$ is real analytic in the complex
$s$-plane; $M(s^*,t)=M(s,t)^*$. 
The right-hand cut starts at $s=4m_{\pi}^2$ while the left-hand 
cut starts at $u=(m_K+m_{\pi})^2$, namely, at $s=-2m_Km_{\pi}+m_{\pi}^2-t\;
(\equiv s_L)$. (See Fig. 1.)  For $t$ fixed small, 
the $s\rightarrow\infty$ limit of $M(s,t)$ is determined 
by Regge asymptotic behavior of $K^*$ exchange. Since  
$\alpha_{K^*}(m_{\pi}^2)\simeq 0.3$, one can write a once-subtracted 
dispersion relation in $s$ at the physical point $t= m_{\pi}^2$ as:
\begin{equation}
   M(s,t) = M(s_0,t) +\frac{s-s_0}{\pi}
 \int_{4m_{\pi}^2}^{\infty}\frac{{\rm Im}M(s',t)}{(s'-s)(s'-s_0)}ds'+
       \frac{s-s_0}{\pi}\int_{-\infty}^{s_L}
       \frac{{\rm Im}M(s',t)}{(s'-s)(s'-s_0)}ds'. \label{dispersion}
\end{equation} 
The decay amplitude is obtained by setting $p_s^{\mu}=0$, namely, 
$s=m_K^2$ and $t= u =m_{\pi}^2$. The first integral in 
Eq. (\ref{dispersion}) due to the $s$-channel 
intermediate states describes the FSI, both elastic and inelastic. 
Its discontinuity across the cut is given by
\begin{equation}
  {\rm Im}M(s,t)_R =\sum_n\pi(2\pi)^3\delta(p_a+p_b-p_n)
\langle\pi|j_{\pi}(0)|n^{out}\rangle \langle n^{out}|H_w(0)|K\rangle.
\end{equation}
The lowest two-body inelastic intermediate state is $K\overline{K}$. 
The triangular diagram depicted in Fig. 2a is the simplest and probably
the most important of the inelastic FSI.
In contrast, the second integral of Eq. (\ref{dispersion}) arises from 
the $u$-channel intermediate states which give the discontinuity, 
\begin{equation}
    {\rm Im}M(s,t)_L =-\sum_n\pi(2\pi)^3\delta(p_K-p_b-p_{n})
\langle\pi|H_W(0)|n^{out}\rangle\langle n^{out}|j_{\pi}(0)|K\rangle.
\end{equation}
The nearest left-hand singularity is generated by the process, $K\rightarrow
\rho K\rightarrow\pi\pi$, namely, a virtual dissociation followed by
a weak decay. (See Fig. 2b.) The diagram of Fig. 2b has no $s$-channel 
singularity and therefore cannot be identified with the FSI. 
Nonetheless, it contributes to 
the $s$ dependence of the amplitude. The process of the $K^*\pi$ 
intermediate state of Fig. 2c belongs to the same class.
We call this class of diagrams as the ISI diagrams. Most generally, a 
single diagram can contribute to both the ISI and the FSI. It is clear 
from the two diagrams in Fig. 2 that the elastic and inelastic FSI 
alone, namely the $s$-channel singularities, are not sufficient to
determine the $s$ dependence of $M(s)$.  It is also obvious that the
phase due to the diagrams of Fig. 2b and 2c are in no way related to
scattering phase shifts of strong interaction at the corresponding energy.

\subsection{Phase representation}
 
  Having defined the ISI by the spurion dispersion relation, we now
write the OM dispersion relation by taking the initial kaon 
off mass shell. If we follow the reduction formula\cite{LSZ}, we would 
reduce one of the pions in addition to the kaon. Rather than handling 
product of three operators $j_{\pi}$, $j_K$, and $H_W$, we may take 
more an intuitive approach by relying on perturbative diagram analysis 
of analyticity\cite{Eden}. 

Choosing a suitable extrapolation field for the kaon, we can write 
a dispersion relation in the variable $p_K^2$ keeping $p_s^{\mu} = 0$. 
Since $s=(p_K+p_s)^2 = p_K^2$ for $p_s^{\mu}= 0$,
 we denote $p_K^2$ by $s$. 
In terms of the Mandelstam variables previously defined, we fix $t$ and 
$u$ to $m_{\pi}^2$ keeping $s$ as the only variable. Then all 
singularities appear on the positive real axis of $s$. (See Fig. 3.) 
The intermediate states $K\overline{K}$, $\rho K$, and $K^*\pi$ of the 
diagrams in Fig. 2 generate cuts starting at $s=4m_K^2$, 
$(m_{\rho}+m_K)^2$, and $(m_{K^*}+m_{\pi})^2$, respectively. To obtain 
the phase dispersion relation of Omn\`{e}s-Muskhelishvili\cite{OM}, 
we write a dispersion relation for the logarithm of the decay amplitude 
$M(s)$. Keeping in mind that zeros of $M(s)$ are singularities of $\ln M(s)$,
we obtain
\begin{equation}
  M(s) = \prod_{i=1}^N\biggl(\frac{s-s_i}{s_0-s_i}\biggr)M(s_0)
         \exp\biggl(\frac{s-s_0}{\pi}\int_{4m_{\pi}^2}^{\infty}
         \frac{\delta(s')}{(s'-s)(s'-s_0)}ds'\biggr), \label{OM}
\end{equation}     
where $\delta(s)=\arg M(s)$. In writing the dispersion integral for 
$\delta(s)$ with only one subtraction, we make the mild assumption 
that $\delta(s)$ does not keep rising or falling indefinitely for large 
$s$. Though the absorptive parts of the ISI diagrams and the inelastic 
FSI diagrams both appear on the right-hand cuts of the OM representation,
they describe physically different processes. When we calculate 
the on-shell decay amplitudes, we are incorporating part of the ISI 
effect through the $K\rightarrow\pi$ form factor. To our knowledge, 
however, nobody has included the ISI process in the OM representation.

\section{Zeros of decay amplitude}

     The large $s$ behavior of $M(s)$ in Eq. (\ref{OM}) is determined by
the number of zeros and the asymptotic value of the phase $\delta(\infty)$. 
It is straightforward to find that, up to logarithmic factors,
\begin{equation}
    \lim_{s\rightarrow\infty}M(s) \sim s^{N-\delta(\infty)/\pi}, 
            \label{asym1}
\end{equation}
where we have normalized as $\delta(4m_{\pi}^2)=0$ at the threshold. 
Therefore, once we know the large $s$ behavior of $|M(s)|$, 
the number of zeros is determined by $\delta(\infty)$.
In potential theory, $M(s)$ is proportional to the inverse Jost function 
$1/f_+(k)$\cite{Fermi}. The Jost function $f_+(k)$\cite{Jost} is analytic in 
the upper half plane of the complex variable $k(=(s/4-m_{\pi}^2)^{1/2})$ 
and approaches unity at $|k|=\infty$. The zeros of $f_+(k)$ mean
bound states. If we repeat the argument leading to Eq. (\ref{asym1}) 
for $f_+(k)$ and use $\arg f_+(k)= -\delta(k)$ and $f_+(\infty)=1$, 
Levinson's theorem\cite{Levinson} results;
           $N_b +\frac{1}{\pi}\delta(\infty)=0$,
where $N_b$ is the number of zeros of $f_+(k)$, {\it i.e.,}
that of bound states.\footnote{Levinson's theorem is usually quoted 
as $N_b = \bigl(\delta(0)-\delta(\infty)\bigr)/\pi$.} In potential theory, 
therefore, $\delta(\infty)=0$ in any channel that has no bound state.  
When inelastic scattering is included, however, $M(s)$ is no longer 
as simple as the inverse of the Jost function of potential 
scattering and therefore Levinson's theorem does not hold. 

It should be emphasized that above the inelastic threshold,
the phase $\delta(s)$ of $M(s)$ has nothing to do with the phase shift
of $\pi\pi\rightarrow\pi\pi$ scattering at the corresponding energy.
The phase of $M(s)$ is a result of complicated interplay of strong and 
weak interactions unlike the scattering phase shift of
strong interaction. (See Appendix A.) Therefore, it is nearly hopeless to 
determine $\delta(\infty)$ either experimentally or directly from theory. 
All we can hope is to determine $N-\delta(\infty)/\pi$, not $N$ 
and $\delta(\infty)$ separately, from the large $s$ behavior of
$|M(s)|$ using Eq. (\ref{asym1}).    

We know that $M(s)$ has one zero in the low-energy region. In the flavor 
SU(3) symmetry limit of strong interaction with no electromagnetic or 
quark mass difference correction, charge-conjugation oddness of 
parity-violating $H_W$ requires that $K\rightarrow\pi\pi$ 
be forbidden in the $I=0$ channel\cite{Gell-Mann}. 
Therefore, the $K\rightarrow\pi\pi$ decay amplitude 
vanishes at $p_K^2=p_a^2=p_b^2$ in this limit.
It was shown in the soft-pion analysis of the 1960's
that the $\Delta I = \frac{1}{2}$ amplitude of $K\rightarrow\pi\pi$ 
vanishes like $\sim 2p_K^2-p_a^2-p_b^2$\cite{softK}.
If SU(3) breaking enters only through the external 
meson momenta, therefore, $M(s)$ should vanish at $s=m_{\pi}^2$. 
This low-energy off-shell behavior is very robust
for the $\Delta I = \frac{1}{2}$ amplitude. The $\Delta I=\frac{3}{2}$ 
decay through pure weak interaction ({\it i.e.,} no electromagnetic
or quark mass difference correction) shows the same behavior.  
However, we do not know how many more zeros $M(s)$ 
may have outside the soft-meson region. To study more about zeros, 
we next look into the large $s$ behavior of $M(s)$.

\section{Extrapolated field for kaon and asymptotic behaviors}

 A well-known theorem of field theory\cite{Kame} states that no matter
what local operator may be used for a particle field, the on-shell S-matrix
is unique after correct wave-function renormalization is made. The flip
side of this theorem is that off-shell matrix elements depend on choice
of a particle field. We obtain the same on-shell low-energy $\pi\pi$ 
scattering whether we may use the linear $\sigma$-model
or the nonlinear $\sigma$-model. Within the nonlinear model, there are
different realizations such as the original version by
Schwinger\cite{Schwinger}, the Callan-Coleman-Wess-Zumino (CCWZ) 
realization\cite{CCWZ}, and so forth\cite{Bando}. All give Weinberg's 
$\pi\pi$ scattering lengths\cite{Weinberg} correctly. In general, 
however, their off-shell behaviors are different.\footnote{
We can see the situation when we express the pion field
$\mbox{\boldmath$\phi$}$ of the linear $\sigma$-model in terms of
the pion field $\mbox{\boldmath$\pi$}$ of the nonlinear $\sigma$-model of
CCWZ: $\mbox{\boldmath $\phi$}= \sqrt{Z_3}(\mbox{\boldmath $\pi$}
-\frac{1}{3f_{\pi}^2}(\mbox{\boldmath $\pi$}\cdot\mbox{\boldmath $\pi$})
+ \cdots)$. The off-shell pion amplitude of the linear $\sigma$-model
is sum of the off-shell one-pion amplitude, the off-shell three-pion
amplitude and so forth of the nonlinear $\sigma$-model. Only when
the pion is on mass shell, do the amplitudes of two models agree with
each other. We encounter the same situation when the kaon field 
is taken off mass shell.}
In the low-energy expansion, symmetry and kinematical constraints
often mask or obscure differences among off-shell amplitudes. 
Nevertheless, the $K\rightarrow\pi\pi$ amplitude 
would depend on choice of the extrapolation field if we included 
the terms of the order higher than $O(p^2)$ in the expansion. 

 S-matrix theory of the 1960's was an attempt to build particle 
physics theory only with on-shell quantities, avoiding off-shell
ambiguities. When we used the 
OM-representation to formulate the FSI, however, we actually stepped over
the premise of S-matrix theory. Unlike the electromagnetic form factors,
to which the OM-representation was most successfully applied\cite{Drell},
we must take a hadron (the kaon in our case) off mass shell in order to
write the OM representation for the $K\rightarrow\pi\pi$ amplitude.
In the case of the electromagnetic form factors, the current operator is
defined at the beginning. Similarly we must first determine what operator 
is used to describe the off-shell kaon.
Only after we have defined the kaon field and have chosen QCD as the 
underlying fundamental theory, do the number and the location of zeros
as well as the asymptotic value $\delta(\infty)$ become unique. 
If one chooses one of the nonlinear realizations and works with 
its effective Lagrangians, the large $s$ behavior of $M(s)$ 
would be far more singular and actually meaningless since their 
applicability is limited to low energies. The amplitude $M(s)$ has 
no unique physical meaning nor direct connection to reality except 
at $s=m_K^2$. The phase theorem assures that only the phase of
the decay amplitude in the elastic rescattering region is unique and
physical. Nonuniqueness of off-shell quantities should cancel out in 
the final result of computation of the physical quantity $M(m_K^2)$, 
but only in principle. We may phrase that the FSI theory formulated with 
the OM representation is not S-matrix theory but field theory even though
it uses the technique of S-matrix theory.

 Let us study the asymptotic behavior in QCD at large $s$, that is,
at $s=O(\mu^2)$ which we assume to be the lower end of the asymptopia. 
Use of the PCAC relation
$\partial_{\lambda}A^{\lambda}=m_P^2f_P\phi$ for the pseudoscalar mesons is
appropriate since we are able to work in perturbative QCD while
maintaining chiral symmetry off shell. 
We obtain the large $s$ limit of $M(s)$ from
\begin{equation}
 -i\frac{G_F}{\sqrt{2}}\biggl(\frac{1}{m_K^2f_K}\biggr)
           \int d^4x e^{i(p_a+ip_b)x}(\Box-m_K^2)
           \langle\pi(p_a)\pi(p_b)|{\rm T}(c_i(\mu){\cal O}_i(0)
           \partial_{\lambda}A^{\lambda}(x))|0\rangle.
\end{equation}
We follow the reasoning of perturbative QCD\cite{Brodsky} 
in the approximation of $m_{u,d}(\ll m_s)\rightarrow 0$. 
Since $\partial_{\lambda}A^{\lambda}=im_s(\overline{q_L}s_R -
\overline{s_R}q_L)$ $(q=u,d)$, one power of $m_s$ appears from 
$\partial_{\lambda}A^{\lambda}$. Furthermore, one insertion of $m_s$ 
is needed for the right-handed $s_R$ (or $\overline{s_R}$) of
$\overline{q_L}s_R - \overline{s_R}q_L$ to turn into $s_L$ 
(or $\overline{s_L}$) of the weak current. Therefore $M(s)$ must
contain $m_s^2$ at large $s$.\footnote{ 
Though the SU(3) breaking due to one power of $m_s$ is almost 
compensated by $1/m_K^2$, the other power of $m_s$ remains as
a large SU(3) breaking. When we use the PCAC for the kaon field, 
therefore, the amplitude extrapolated to the large $s$ limit 
shows a large SU(3) breaking through the current quark mass difference.}

Next we look into chirality matching of the $u,d$ quarks. We first examine 
the case that ${\cal O}_i=J_{\mu}^{\dagger}J^{\mu}$ consists of left-handed 
currents alone. The leading quark diagram is shown in Fig. 4a. 
In this case, the factor $f_{\pi}p^{\mu}_{\pi}$ arises from the 
formation of each pion (the oval blobs in Fig. 4).
Together with $m_s^2$, therefore, we obtain the large $s$ behavior of
the form
\begin{equation}
     M(s)\sim \frac{G_Fm_s^2f_{\pi}^2 s}{m_K^2f_K}
           \label{asym2}
\end{equation}
up to a power of $(\ln s)^{-1}$.
In contrast, when ${\cal O}_i$ consists of left and right-handed 
currents as the dominant penguin operator ${\cal O}_6$ does, 
the pion is formed from a chirality-matched pair of $\overline{q_L}q_R$ 
or $\overline{q_R}q_L$ with the matrix element $m_{\pi}^2f_{\pi}/(m_u+m_d)$.
This type of decay dominates at low energies, 
{\it e.g.,} on the kaon mass shell\cite{Buchalla}.  
When we go to high energies, however, the absence of $p^{\mu}$ in 
the pion formation makes this penguin amplitude softer in $s$ dependence:
\begin{equation} 
     M(s)\sim \frac{G_Fm_s^2f_{\pi}^2 m_{\pi}^4}{m_K^2f_K(m_u+m_d)^2},
    \;\;({\rm for}\;{\cal O}_6).          \label{asym3}
\end{equation}
When we combine Eq. (\ref{asym2}) or (\ref{asym3}) with Eq. (\ref{asym1}), 
we obtain $N - \delta(\infty)/\pi = 1$ or 0.
{\it If} the low-energy zero at $s=m_{\pi}^2$ is the only zero of
$M(s)$, then the asymptotic phase should approach $0$ or $\pi$,
depending on the asymptotic behavior of $M(s)$: 
\begin{eqnarray}
      \delta_0(\infty)= \left\{ \begin{array}{cl} 
                        0 &(N=1;\; M(s)\sim s), \\
                       \pi&(N=1;\; M(s)\rightarrow{\rm const}).
                  \end{array} \right.      \label{asym5}
\end{eqnarray}
The first line applies to the ``tree'' operator ${\cal O}_{1,2}$,
while the second line is for the dominant penguin operator ${\cal O}_6$.
We emphasize that these asymptotic behaviors are specific
to the case of the PCAC in QCD. They are not necessarily applicable
to effective theories for which we do not know $M(\infty)$. 

For the $\Delta I = \frac{3}{2}$ decay, the pure weak decay 
has the zero at $s=m^2_{\pi}$. (Appendix B) If this is the 
only zero ($N=1$), we have $\delta_2(\infty)=0$ since $M(s)\sim s$.
On the other hand the EM penguin matrix element need not vanish 
at $s=m_{\pi}^2$. The perturbative QCD argument presented 
above remains valid even after one internal photon line is inserted.
Therefore $M(s)\sim constant$ for the dominant penguin operator 
${\cal O}_8$. The cases of the smallest number of zeros 
for the $\Delta I = \frac{3}{2}$ amplitudes are, therefore,
\begin{equation}
        \delta_2(\infty) =  \left\{ \begin{array}{cl}
           0&\;(N=1;\;M(s)\sim s) \\
                0&\;(N=0;\;M(s)\sim\;{\rm constant}) ,\\
                -\pi&\;(N=0;M(s)\sim s).
                \end{array} \right.  \label{asym6}
\end{equation}
The first line applies to the pure weak decay ${\cal O}_{1,2}$, 
while the second line is for the EM penguin ${\cal O}_8$.

If there are more zeros in the amplitude, $\delta_I(\infty)$ shifts 
upward from those given in Eqs. (\ref{asym5}) and (\ref{asym6}). 
The larger the value of $|\delta_I(\infty)|$ is, the more important
the contributions from the inelastic FSI and the ISI are, as we 
shall see below. 
 
\section{Quantitative discussion}

We first define the problem to which we address in this paper: Suppose 
that we compute with the effective weak interaction $c_i(\mu){\cal O}_i$
the decay amplitude $M(s)$ at some small value $s_0$ below the $\pi\pi$
threshold. The value of such $M(s_0)$ is real though it incorporates all 
the FSI and the ISI that enter through the dispersion integral of 
$\delta(s)$. Then we ask how much $|M(m_K^2)|$ is different from 
$M(s_0)$. We should keep in mind that the value of $M(m_{\pi}^2)$ 
depends on choice of the kaon field in principle. Strictly speaking,
therefore, this problem makes sense only after we have defined the 
kaon field.  

We focus on the least ambiguous case that $M(s)$ 
has the minimum number of zeros, namely, only one zero at 
$s=m_{\pi}^2$ for all amplitudes except for the EM penguin and 
no zero for the EM penguin. The zero at 
or near $s=m_{\pi}^2$ is robust. In contrast, it is rather 
an ad hoc assumption that there is no other zero. We use the values 
of $\delta_I(\infty)$ predicted by the PCAC for the kaon, namely, 
$\delta_0(\infty)=\pi$ for the dominant gluon penguin operator
of left and right-handed currents 
and $\delta_2(\infty)=0$ for the $\Delta I=\frac{3}{2}$ operators. (cf Eqs. 
(\ref{asym5}) and (\ref{asym6}).) 

When $H_W$ is one of the effective operators $c_i(\mu){\cal O}_i$, it is 
reasonable to match the OM representation to the short-distance QCD by 
cutting off the dispersion integral at $s\simeq\mu^2$. Since $\mu^2$ is 
the spacelike scale of renormalization group, it would be more appropriate 
to make analytic continuation of $c_i(\mu)$ to the timelike region.
This would generate an imaginary part of $O((\ln\mu)^{-1})$ relative
to the real part as the high-energy tail of FSI. We can ignore it 
as a small correction if $\mu$ is in the perturbative regime of QCD. 
It has been a subject of discussion how far one can lower $\mu$ without 
losing desired accuracy and how one should match $c_i(\mu)$ to 
low-energy matrix element computation. If we could choose as low as 
$\mu=$ 1 GeV, the issues of this paper would mostly disappear since all we 
would need is the elastic scattering phase shifts.  However, such a low 
cutoff is hard to justify particularly in the $I=0$ channel where
the scalar resonances $f_0$ exist at 1500 MeV and 1700 MeV. 
While $\mu\simeq$ 2 GeV is probably safe, it may be possible 
to lower $\mu$ close to 1.5 GeV ($\simeq m_c$). Near $s=\mu^2$, 
$\delta(s)$ ought to approach its asymptotic value $\delta(\infty)$. 
Otherwise the $s$ dependence of $M(s)$ computed with the cutoff dispersion 
integral would not give a smooth asymptotic behavior at $s=\mu^2$. 

Below the inelastic threshold, the phase of $M(s)$ 
is equal to the $s$-wave phase shift of elastic $\pi\pi$ scattering by
Watson's theorem\cite{Fermi}. The elastic phase shift of $I=0$ is 
positive and grows to large values with energy\cite{pipi}, while that 
of $I=2$ stays negative and small in magnitude. Though theoretically 
the inelastic scattering starts at $\sqrt{s}=4m_{\pi}$, in reality 
the inelasticity in the $I=0$ channel appears rather suddenly at 
$\sqrt{s}=2m_K$ where the $K\overline{K}$ channel opens. The $I=0$ 
scattering phase shift rapidly rises to $\pi$ across $\sqrt{s}=2m_K$ 
passing through the $f_0(980)$ resonance just below the $K\overline{K}$
threshold. As we have emphasized, however, the $\pi\pi$ scattering 
phase shift above $\sqrt{s}=2m_K$ has little to do with the phase of 
the $K\rightarrow\pi\pi$ decay amplitude: Once an inelastic channel 
opens, the phase theorem no longer relates the two phases since the 
latter depends on the relative sign and magnitude of the $K\rightarrow 
K\overline{K}$ transition matrix element to the $K\rightarrow\pi\pi$ decay 
amplitude too, as we shall see below. Therefore the surge of the $\pi\pi$ 
scattering phase shift across $\sqrt{s}=2m_K$ does not necessarily means 
that the phase $\delta_0(s)$ of $M(s)$ rises abruptly in the same way. 
In fact, when an elastic channel strongly couples to an inelastic channel, 
a partial wave amplitude moves quickly toward the center of the 
Argand diagram and its scattering phase shift often undergoes a rapid 
variation across the inelastic threshold. It is quite possible that
the phase $\delta_0(s)$ of $M(s)$ varies more slowly across 
$\sqrt{s}=2m_K$ than the $\pi\pi$ scattering phase shift does.

With the asymptotic values of $\delta_0(\infty)=\pi$ and 
$\delta_2(\infty)=0$ (cf Eqs. (\ref{asym5}) and (\ref{asym6}), 
we have drawn schematically 
the behaviors of the phase $\delta_I(s)$ of $M(s)$ up to 
$s=\mu$($\simeq$ 2 GeV) in Fig. 5a 
by smoothly interpolating between $\sqrt{s}=2m_K$ and $\mu$.

\subsection{Inelastic FSI and ISI}
 
The phase $\delta_I(s)$ above the inelastic threshold comes from the ISI 
and the inelastic FSI. For curiosity, we have actually computed the 
diagrams that generate the lowest singularities in the OM representation
(Fig. 2a and 2c). The FSI contribution of the diagram in Fig. 2a 
to the absorptive part ${\rm Im}M(s)$ of $I_{\pi\pi}=0$ is:
\begin{equation}
     {\rm Im}M^{K\overline{K}}(s) 
 = \biggl(\frac{g^2_{K^*K\pi}}{4\pi}\biggr)
 \sqrt{\frac{s-4m_K^2}{s}}I(s)\kappa(s){\rm Re}M(s),
\end{equation}
where $\kappa(s)$ is the ratio of the $K\rightarrow K\overline{K}$ 
amplitude of ${\Delta I=1/2}$ to the $K\rightarrow\pi\pi$ amplitude of 
${\Delta I=1/2}$ at $\sqrt{s}$, $g^2_{K^*K\pi}/4\pi\simeq 1.3$ 
is the $K^*K\pi$ coupling, and $I(s)$ is a known logarithmic function 
of $\sqrt{s}$. We have chosen $\frac{1}{2}\sum(|out\rangle\langle out|)+ 
|in\rangle\langle in|)$ for the intermediate states  
in order to ensure the absorptive part to be real. For a numerical 
estimate of the magnitude, let us choose some value for $\sqrt{s}$, 
say, $\sqrt{s}=2.5m_K (\simeq 1.25$ GeV), for which $I(s)\simeq 0.50$. 
We set $\kappa(s)=(s-m_K^2)/(s-m_{\pi}^2)$, which is the prediction
of the octet dominance for $H_W$ and SU(3) symmetry for strong 
interactions. Then we find
\begin{equation}
       {\rm Im}M^{K\overline{K}}\simeq 0.39\times {\rm Re}M
          \;\;({\rm at}\;\sqrt{s}= 2.5 m_K).
\end{equation}
At higher $\sqrt{s}$, the states such as $\eta\eta$, $a_1\pi$, 
and $\rho\rho$ become important intermediate states. While reliable 
estimate is less easy for those decay modes, the inelastic FSI of 
$K\rightarrow K\overline{K}\rightarrow\pi\pi$ appears sizable.

Let us turn to the ISI of the diagram of Fig. 2c which generates the
lowest ISI singularity. There are two independent 
$\Delta I = \frac{1}{2}$ transitions for $K^*\rightarrow \rho\pi$:
\begin{equation}
  a(\mbox{\boldmath$\pi$}\cdot\mbox{\boldmath$\rho$})
   (\overline{K}S)
 +b(\mbox{\boldmath$\pi$}\times\mbox{\boldmath$\rho$})\cdot
   (\overline{K}\mbox{\boldmath$\tau$}S) + h.c.,  \label{3/2}
\end{equation}
where $S=(0,1)$ represents the $\Delta I=\frac{1}{2}$ spurion. 
The first term of Eq. (\ref{3/2}) cancels out in the diagram of Fig. 2c 
by the isospin structure. The second term gives   
\begin{equation}
{\rm Im}M^{K^*\pi}(s) = 
   \biggl(\frac{g_{K*K\pi}g_{\rho\pi\pi}}{4\pi}\biggr)
 \bigg(\frac{2|{\bf q}|}{\sqrt{s}}\biggr)\lambda(s)J(s){\rm Re}M(s),
\end{equation}
where $\lambda(s)$ is the ratio $b/{\rm Re}M(s)$, ${\bf q}$ is the
{\it c.m.} momentum of $K^*\pi$, and $J(s)$ is
another known function of $\sqrt{s}$.
In contrast to the case of the inelastic FSI of the $K\overline{K}$
intermediate state above, we know little about the value of 
$\lambda(s)$. For $\lambda(s)\approx 1$ we obtain at $\sqrt{s}=2.5m_K$
($J(s)\simeq 0.36$)
\begin{equation}
   {\rm Im}M_0^{K^*\pi}\approx  0.27 \times {\rm Re} M_0
          \;\;({\rm at}\; \sqrt{s} = 2.5 m_K), \label{ISI}
\end{equation} 
The ISI is not negligible either though Eq. (\ref{ISI}) gives only 
an order-of-magnitude estimate at best. The intermediate states $\rho K$, 
$\omega K$, and so forth start contributing to the ISI at a little above 
$\sqrt{s}= 2.5m_K$. 

This exploratory numerical exercise indicates that both the ISI 
and the inelastic FSI are large enough to deserve study.
Since computation by individual diagrams involves too many uncertainties, 
we instead attempt to extract general trends and some quantitative
results with minimal assumptions. 

\subsection{$I=0$ channel}

We estimate the variation $M(m_K^2)-M(s_0)$ for the dominant 
gluon penguin decay with $\delta_0(s)$ of Fig. 5a.  
We choose $s_0$ at the SU(3)-symmetry point $m_{\pi}^2$ and quote 
the variation due to the exponential in the $\pi\pi$ isospin
channel $I$ after separating out the factor $(s-m_{\pi}^2)$:
\begin{eqnarray}
 M_I(s) &\equiv& (s-m_{\pi}^2)\tilde{M}_I(s) \nonumber \\     
{\cal R}_I &=& |\tilde{M}_I(m_K^2)|/\tilde{M}_I(m_{\pi}^2).
\end{eqnarray}
The result is
\begin{equation}
     {\cal R}_0  \simeq \left\{ \begin{array}{cl}
                  1.50 &(\mu = 1.5\;{\rm GeV}), \\
                  1.57 &(\mu = 2\;{\rm GeV})
                 \end{array} \right.  \label{N1}
\end{equation} 
The subtraction in the dispersion integral suppresses the contribution 
from the high-energy tail at 1.5 GeV $< \sqrt{s} <$ 2 GeV.
The authors of Ref.\cite{Pich} quoted ${\cal R}_0 = 1.41\pm 0.06$ without
including the contribution above the inelastic threshold.
We can estimate the effect of the ISI 
and the inelastic FSI by evaluating the phase integral above 
$\sqrt{s}=2m_K$ separately from the elastic region $\sqrt{s}\leq 2m_K$. 
The contribution from the ISI and the inelastic FSI is:
\begin{equation}
       {\cal R}_0^{ISI+IFSI} = \left\{  \begin{array}{cl}
                 1.13 &(\mu = 1.5 \;{\rm GeV}),  \\
                 1.18 &(\mu = 2 \;{\rm GeV}). 
               \end{array}  \right. \label{N2}
\end{equation}
The ISI and the inelastic FSI combined contribute to the variation of 
$M_0(s)$ by a third to one half of the elastic FSI. The sizable
contribution above the inelastic threshold means that we cannot truncate
the integral at $\mu=$ 1 GeV and connect to short-distance calculations. 
This is an unfortunate situation since we cannot obtain  
the values of $\delta_0(s)$ at $2m_K <\sqrt{s} <\mu$ directly
from experiment. For comparison we have made an estimate in 
the case that $\delta_0(s)$ follows the behavior of the $\pi\pi$
phase shift across the $K\overline{K}$ threshold. (see the broken 
curve in Fig. 5a.) The value of ${\cal R}_0^{ISI+IFSI}$ in this case is:
\begin{equation}     
   {\cal R}_0^{ISI + IFSI} = \left \{ \begin{array}{cl}
                 1.17 &(\mu = 1.5\;{\rm GeV}), \\
                 1.23 &(\mu = 2 \;{\rm GeV}).
               \end{array}  \right.   \label{N3}
\end{equation}  
Though the ISI and the inelastic FSI are important, their magnitudes are not
highly sensitive to the $s$ dependence of $\delta_0(s)$. As we shall see
below, it is the value of $\delta_0(\infty)$, namely, the number of zeros 
of the amplitude that they are more sensitive to.

\subsection{I=2 channel}

We consider the case of $\delta_2(\infty)=0$ for the pure weak decay
($N=1$) and the EM penguin decay ($N=0$).
Because of the large enhancement of the $\Delta I = \frac{1}{2}$
decay over the $\Delta I=\frac{3}{2}$ decay, the $I = 2$ channel of 
$K_S$ may receive a sizable $\Delta I = \frac{3}{2}$ 
contribution from an enhanced $\Delta I=\frac{1}{2}$ decay 
followed or preceded by long-distance electromagnetic isospin
breaking, {\it e.g.,} $\pi$-$\eta$ mixing.\footnote{In the papers where
Cabibbo and Gell-Mann\cite{Gell-Mann} proved the theorem of the
vanishing $K\rightarrow\pi\pi$ amplitude of $\Delta I=\frac{1}{2}$,
they even speculated on the possibility that the $\Delta I = 3/2$
decays could be attributed entirely to the electromagnetic correction
to the $\Delta I = 1/2$ transition. The long-distance contributions 
from $\pi-\eta$, $\pi-\eta'$, and some intermediate states 
containing a photon were studied without FSI\cite{Donoghue}.} 
Such a contribution may compete with the pure weak decay and
the EM penguin decay in the $\Delta I = \frac{3}{2}$ decay.
For the $\Delta I = \frac{3}{2}$ decay generated by long-distance 
electromagnetic corrections, however, $\delta_2(s)$ of $M_2(s)$ is 
no longer equal to the $\pi\pi$ scattering phase shift of $I=2$ 
nor $I=0$ even below $\sqrt{s}=2m_K$.

When we ignore the long-distance electromagnetic correction 
and use $\delta_2(s)$ of Fig. 5a, the numbers 
corresponding to Eqs. (\ref{N1}) and (\ref{N2}) are:
\begin{equation}
       {\cal R}_2 =  0.91 \;(\mu = 1.5\;{\rm GeV}), \label{I=2}
\end{equation}
and
\begin{equation}
       {\cal R}_2^{ISI+IFSI} = 0.99 \;(\mu = 1.5 \;{\rm GeV}). \label{I=2'}
\end{equation}
Our numbers are in perfect agreement with ${\cal R}_2\simeq 0.92\pm 0.02$
obtained in \cite{Pich}. However there is no factor of $(s-m_{\pi}^2)$ for 
the $\Delta I = \frac{3}{2}$ EM penguin decay. The contribution from 
above 1 GeV is much smaller in $I=2$ than in $I=0$ because of 
$\delta_2(\infty)=0$. For the same reason, dependence on the 
value of $\mu$ is very mild. For the EM penguin operators ${\cal O}_{9,10}$,
$M(s)\sim s$ so that the asymptotic phase would be $\delta_2(\infty)
=-\pi$ (cf Eq. (\ref{asym6})). In this case we obtain a 
stronger suppression. It should be emphasized again that Eq. 
(\ref{I=2}) does not apply to long-distance electromagnetic corrections to 
the gluon penguin operator decay, since we cannot equate $\delta_2(s)$ 
of $M_2(s)$ to a $\pi\pi$ scattering phase shift even in the elastic region. 
The OM representation is practically useless in this case.

\subsection{More zeros ($N\geq 2$)}

When $M_0(s)$ has more than one zero, the asymptotic value 
$\delta_0(\infty)$ is larger. Consider the case that $M_0(s)$ 
has one more zero ($N=2$). In this 
case $\delta_0(\infty)=2\pi$. The second zero must be 
located on the real axis in order to be consistent with real 
analyticity $M(s^*)=M(s)^*$. The zero at $s=m_{\pi}^2$ is obtained in the 
Taylor expansion to $O(p^2)$ with SU(3) symmetry. When we include $O(p^4)$,
the location of this zero stays as long as SU(3) symmetry is maintained, but
another zero emerges. 
If the second zero is outside the soft-meson region, we cannot 
trust its presence. If it is inside or close 
to the soft-meson region, it means that the next-order 
terms are important at low energies. If the lowest-order
chiral Lagrangians give a good description of $K$ decays, the second 
zero should be far from the soft-meson region and therefore 
the second zero factor $(s-s_2)$ would not generate very rapid variations  
at low energies. Our hope is that the dominant penguin operator has no
second zero and reaches its asymptopia quickly.
If the second zero exists in the region of $|s|=O(16\pi^2\Lambda_{QCD}^2)$, 
the variation due to the second zero factor $(s-s_2)$ could be 
substantial when we extrapolate from $s=m_{\pi}^2$ 
to $m_K^2$. Lack of our knowledge of its precise location introduces a
large uncertainty in the extrapolation.

If we assume that $\delta_0(s)$ rises linearly in $\sqrt{s}$ to $2\pi$ 
from 1 GeV to 2 GeV (Fig. 5b), the ratio of the exponential factors 
at $s=m_K^2$ and $m_{\pi}^2$ after removing $(s-m_{\pi}^2)(s-s_2)$ is:
\begin{equation}
    {\cal R}_0 = 1.64 \;(\mu = 2\;{\rm GeV}).
\end{equation}
In comparison, if $\delta_0(s)$ follows the inelastic behavior of 
$\pi\pi$ scattering phase shift (the broken curve in Fig. 5b), the
corresponding value could be:
\begin{equation} 
    {\cal R}_0 = 1.83 \; (\mu = 2\;{\rm GeV}).
\end{equation}
In the case of two zeros, therefore, the ISI and inelastic FSI contribution 
can be as large as or even larger than the elastic FSI contribution.

If $M(s)$ has three zeros, the second and third zeros can appear
off the real axis in a complex conjugate pair. 
After we include the variation due to the factor $\prod_i(s-s_i)$,
the uncertainties get quickly out of control as $N$ increases.   
The uncertainty due to the number of zeros and their locations 
appears to be by far the largest.

\section{Conclusion}
 
We have studied with the Omn\`{e}s-Muskhelishvili representation
the off-shell kaon mass dependence of the $K\rightarrow\pi\pi$ decay 
amplitudes for the purpose of exposing the uncertainties that have so 
far not been seriously studied. Determining the phases
above the inelastic threshold encounters two difficulties. One is
complexity of inelastic rescattering and the other is physical nonuniqueness
of off-shell amplitudes in general. In this paper we have studied 
the off-shell behavior with QCD as the underlying fundamental theory
and with the PCAC for the extrapolated kaon field. As it  
is suspected, the largest uncertainty is how many zeros exist
in the amplitudes and where they are located. We regret that we
cannot be optimistic in our conclusion as to accuracy or certainty of
the off-shell extrapolation. If we work in the 
spurion scattering formalism where all hadrons are on shell, we do not 
encounter the physical nonuniquness due to off-shellness. 
However, evaluation of the dispersion relation is equally formidable since
it requires knowledge of the hadron-spurion scattering with the spurion 
carrying nonvanishing energy-momentum. In either approach, we are
unable to work only with physical quantities directly 
measurable in experiment. If we were able to lower 
the cutoff to $\mu =$ 1 GeV, we could do without the 
inelastic region.  However, this does not seem possible since 
$\delta_0(s)$ is not yet asymptotic at 1 GeV. 

If we have very accurate knowledge of $M(s)$ in the low-energy region,
we can suppress the large $s$ portion of the integral by writing
a phase dispersion integral with more than one low-energy 
subtraction\cite{Pich}. 
In this case the low-energy values of $M(s)$ that we feed in the dispersion
relation would incorporate implicitly the contribution from inelastic 
region. 
The problem would then reduce to determining the physically unique value 
of $M(m_K^2)$ with the elastic phase shift and with the off-shell values 
of $|M(s)|$ which are well defined only after field theory is specified for
mesons. Though the physical nonuniqueness of the off-shell $|M(s)|$ should, 
in principle, cancel out in $M(m_K^2)$, it is not clear whether 
we can obtain, in practice, accurate enough low-energy values of
$M(s)$ from higher-order chiral Lagrangian terms. 

Our final comment is on the implications of our study in $B$ decay. 
The FSI is highly inelastic on the $B$ mass shell and the ISI starts 
at $s=(m_{B^*}+m_{\pi})^2$, right above the $B$ mass. Unless 
short-distance strong interactions completely dominate in the FSI and 
the ISI, computation of magnitudes of decay amplitudes would be subject 
to very large uncertainties, far more than in $K$ decay. Several 
theoretical arguments have been presented in favor of short-distance 
dominance after factorization. Nevertheless, 
if we should find {\it from experiment} an FSI 
phase much larger than the short-distance prediction in some decay mode, 
our work here would be telling us that it is almost hopeless to compute 
such a decay amplitude with decent accuracy.
   
\acknowledgements

This work was supported in part by the Director, Office of Science, Division
of High Energy and Nuclear Physics, of the U. S. Department of Energy 
under Contract DE-AC03-76SF00098 and in part by the National Science
Foundation under Grant PHY-95-14797.

\appendix
\section{Phase in the presence of inelasticity}

Many of us are aware that the phase of decay amplitude is not determined
by strong interaction alone at energies where 
inelastic rescattering occurs. Its value is an outcome of 
complicated interplay between strong and weak interactions.
Sicne there is some confusion on this important point in literature, 
we clarify it in the simplest possible term here.

Consider the decay amplitude for $A\rightarrow a+b$ at energies where
scattering of $a+b$ is inelastic in general. The partial-wave S-matrix of 
strong interaction scattering is diagonalized by eigenchannels:
\begin{equation}
         \langle n^{out}|n'^{in}\rangle = \delta_{nn'}e^{2i\delta_n}.
\end{equation}
Making time reversal on $\langle n^{out}|H_w|A\rangle$, we obtain for
T-invariant $H_W$
\begin{eqnarray}
   \langle n^{out}|H_W|A\rangle &=& \langle A|H_W|n^{in}\rangle,
                      \nonumber \\
         &=& \langle A|H_W|n^{out}\rangle\langle n^{out}|n^{in}\rangle,
                      \nonumber \\
         &=& e^{2i\delta_n}\langle A|H_W|n^{out}\rangle.
\end{eqnarray}
With $\langle A|H_W|n^{out}\rangle = \langle n^{out}|H_W|A\rangle^*$, 
this relation leads us to
\begin{equation}
            M_n(s) = \overline{M}_n(s)e^{i\delta_n(s)},
\end{equation}
where $\overline{M}_n(s)$ is real. This is the phase theorem
referred to as Watson's theorem or, to be more precise\cite{Gas},
the Watson-Fermi-Aizu theorem. At energies where only one eigenchannel
is open in the final state, the phase of a matrix element for any 
operator having the same quantum numbers as $H_W$ is universal and equal 
to the scattering phase shift.

However, the situation is different at energies where inelastic channels
are open and a final state is not an eigenchannel of scattering.
Because of T-invariance of strong interaction, the transformation 
between $|ab\rangle$ and eigenchannels $|n\rangle$ can be chosen 
as orthogonal:
\begin{equation}
         |ab^{out}\rangle = \sum_n O_{ab,n}|n^{out}\rangle,
         \;\;(O^*_{ab,n}=O_{ab,n}).
\end{equation}
Therefore the decay amplitude $M_{ab}$ into the observable final state 
$|ab\rangle$ takes the form 
\begin{equation}
     M_{ab}(s) =\sum_n O_{ab,n}(s)\overline{M}_n(s)e^{i\delta_n(s)}.
\end{equation}
The net phase of $M_{ab}(s)$ results from a superposition of
the eigenphase shifts weighted with the orthogonal transformation 
matrix elements $O_{ab,n}(s)$ and the weak decay amplitudes 
$\overline{M}_n(s)$ into the eigenchannels. 

In contrast, the {\it phase shift} $\delta_{ab}$ of scattering 
$a+b\rightarrow a+b$ in the presence of inelasticity
is defined with the inelasticity $\eta_{ab}$ by
$\langle ab^{out}|ab^{in}\rangle \equiv 
\eta_{ab} e^{2i\delta_{ab}}$\cite{Pilkuhn}. 
Therefore,
\begin{equation}
       \eta_{ab} e^{2i\delta_{ab}}=\sum_nO_{ab,n}^2 e^{2i\delta_n}.
\end{equation}
The phase $\delta_{ab}$ is a quantity determined by strong interactions 
alone. Above the inelastic threshold the phase of the decay amplitude 
$M(s)_{ab}$ has nothing to do with the strong interaction phase shift.
 
\section{Low-energy zeros}

It is straightforward to extend the theorem by Cabibbo and Gell-Mann 
\cite{Gell-Mann} to the $\Delta I= \frac{3}{2}$ decay of
$K\rightarrow\pi\pi$ and show that the $\Delta I = \frac{3}{2}$ amplitude 
also vanishes in the SU(3) limit of strong interaction.
We reconstruct the extension of the proof and show the robustness
of the low-energy zero. 

When $H_W$ is the current-current interaction in a symmetric product of two
octet $V-A$ currents, it forms $\{8_S\}$ and $\{27\}$. Using the
SU(3) tensor notation (1,2,3) for ($u,d,s$), we can express the
transformation property of $H_W$ as
\begin{equation}
               H_W \sim H^{12}_{31} + H^{31}_{12},
\end{equation}
where the upper and lower pairs of indices are both symmetric 
under interchange.
To remove $\{8_S\}$ from the right-hand side, we should subtract the 
trace portion. But it is not necessary here since the theorem was already
proven for the octet spurion. The relative sign of  $H^{12}_{31}$ and
$H^{31}_{12}$ is positive. In the case of $\{8_S\}$ this relative sign 
aligns $H_W$ along $\lambda_6$ instead of $\lambda_7$. It is crucial
to the proof of the theorem for the ${\{27\}}$ spurion as well.

We parametrize the $0^-\rightarrow 0^-+0^-$ decay matrix elements of 
$\{27\}$ without derivatives in terms of the $3\times 3$ octet 
meson matrix $M$, instead of $U=\exp(i\lambda_a\phi_a/f_{\pi})$ of 
SU(3)$_L\times SU(3)_R$:
\begin{equation}
    {\cal M} = (M^1_iM^i_3M^2_1 +M^i_1M^3_iM^1_2) +(1\leftrightarrow 2)
                + (3\leftrightarrow 1) 
                + (1\leftrightarrow 2,\;3\leftrightarrow 1). \label{B1}
\end{equation}
The relative sign within each bracket is positive, as pointed out above.
Otherwise ${\cal M}$ would point along a wrong component of $\{27\}$.
 
The parity-violating part of $H_W$ is odd under charge conjugation C
when $H_W$ is (approximately) CP invariant. That is, $H^{12}_{31}
\rightarrow -H^{31}_{12}$ under C for $H_W^{{\rm pv}}$.
It means that the $\{27\}$ (and also $\{8_S\}$) made of symmetric products
of the $V$ and $A$ currents must be a ${\cal C}$-odd multiplet of 
SU(3).\footnote{
The SU(3) charge parity ${\cal C}$ is defined for a self-charge-conjugate
multiplet $\{n\}$ by the sign of
$|\{n\};I,I_3,Y\rangle \rightarrow \pm|\{n\};I,-I_3;-Y\rangle$ under C,
or in the tensor notation, $T^{ij\cdots}_{kl\cdots}\rightarrow\pm 
T^{kl\cdots}_{ij\cdots}$.}
Since $M\rightarrow M^T$ ($T$ = transposed) under C, all terms in 
Eq. (\ref{B1}) transform into each other such 
that ${\cal M}\rightarrow +{\cal M}$. Namely, ${\cal M}$ can make only a 
${\cal C}$-even $\{27\}$. Therefore, we cannot construct an SU(3) covariant 
matrix element having the correct charge conjugation property.

It is obvious in the chiral Lagrangian approach that
one cannot write nonderivative Lagrangian terms for $\{8\}_L$ nor
$\{27\}_L$ of $H_W$. However, chiral symmetry is not really needed 
to prove the theorem; the original proof by Cabibbo and Gell-Mann
never used it.

 The theorem fails once SU(3) breaking is included.
The decay spurion arising from $H_W$ with the electromagnetic interaction
or the quark mass difference 
is no longer a symmetrized product of two octets. The EM penguin contains 
$\{10\}$, and $\{\overline{10}\}$ too. 

The thereom is valid even after derivatives are included for the mesons: 
$M\partial_{\mu}M\partial^{\mu}M$ generates $(p_2\cdot p_3)$, which is
equal to $(p_1^2-p_2^2-p_3^2)/2$ and reduces to a common value 
$-p^2/2$ at the SU(3)-symmetry point, $p_1^2=p_2^2=p_3^2=p^2$.
Therefore all contracted pairs of derivatives are factored out as
SU(3) invariants so that
the SU(3) symmetry property of $M\partial_{\mu}M\partial^{\mu}M$ 
is identical with that of $MMM$. 

Given this theorem, the off-shell $K\rightarrow\pi\pi$ amplitude 
to $O(p^2)$ is unique up to a proportionality constant.
To prove it, write the off-shell amplitude as a function of 
$p_a^2$, $p_b^2$, and $p_K^2$ by incorporating the Bose statistics 
for the final $\pi_a\pi_b$ in $s$-wave:
\begin{equation}
    {\cal M} = A(p_a^2+p_b^2) +Bp_K^2 +O(p^4). 
\end{equation}
In order for ${\cal M}$ to vanish at the SU(3) symmetry point, 
$B$ must be equal to $-2A$. Consequently,
${\cal M}=-A(2p_K^2-p_a^2-p_b^2)$. This argument applies to 
both $\{8\}$ and $\{27\}$.

All we need for the proof is $M\rightarrow +M^T$ and the fact that 
$\Delta I=\frac{3}{2}$ of $H_W^{{\rm pv}}$ is a ${\cal C}$-odd $\{27\}$. 
Therefore, even if the extrapolation fields of $0^-$ mesons did not
respect chiral symmetry off mass shell, the theorem would be valid.
It is a misnomer to refer to the zero at $s=m_{\pi}^2$ as 
the ``chiral zero''.

%%%%%%%%%%%%%%%%%%%%%%%%%%%%%%%%%%%%%%%%%%%%%%%%%%%%%%%%%%%%%%%%
%\begin{table}
%\caption{}
%\begin{tabular}{cccc} 
%
%\end{tabular}
%\label{table:1}
%\end{table}
%%%%%%%%%%%%%%%%%%%%%%%%%%%%%%%%%%%%%%%%%%%%%%%%%%%%%%%%%%%%%%%%%

%\input psfig
\noindent
\begin{figure}
\epsfig{file=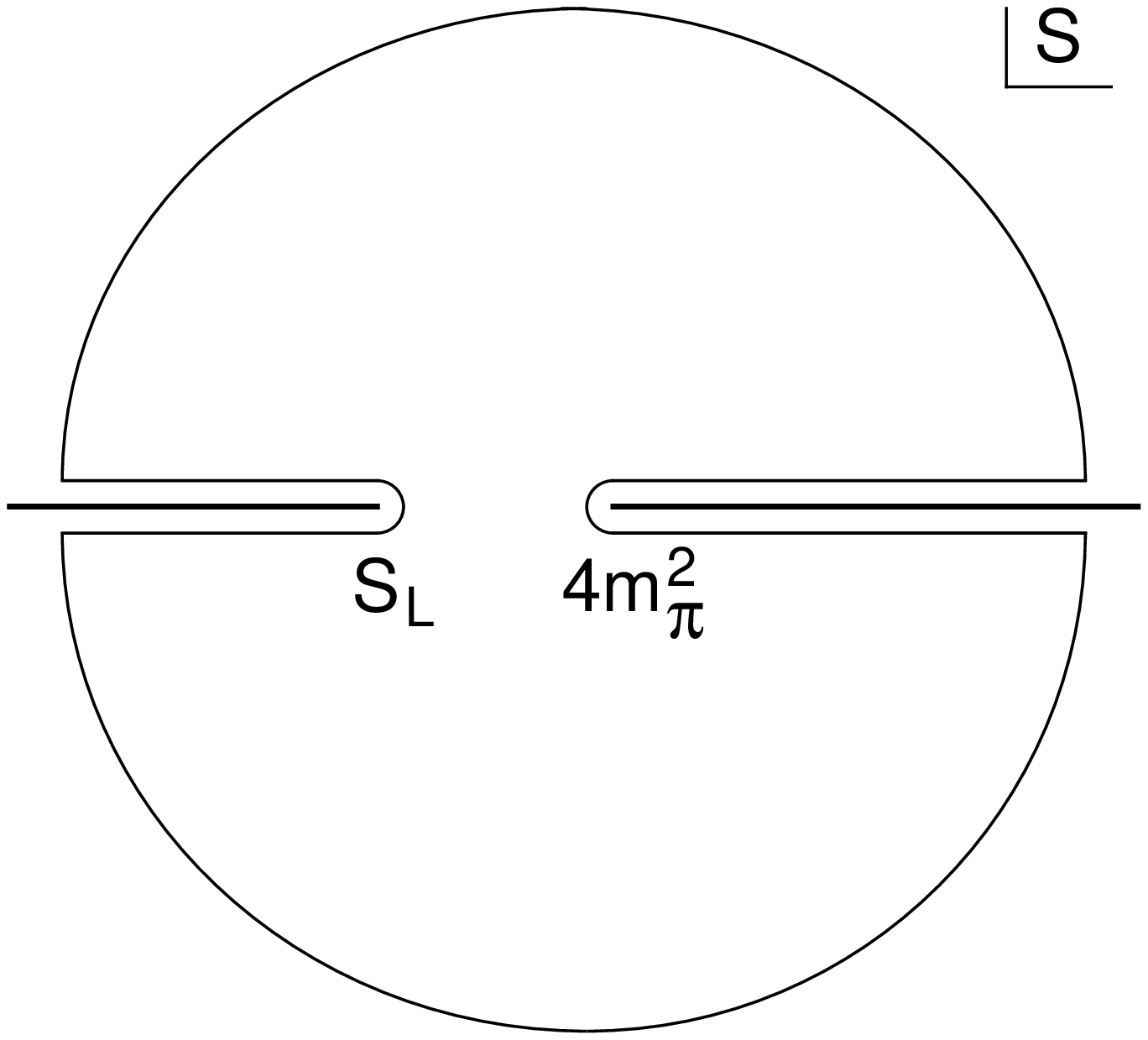,width=5cm,height=4.6cm}
\caption{Analyticity of the spurion scattering amplitude off $K$ 
 into $\pi\pi$. 
\label{fig:1}} 
\end{figure}

\noindent
\begin{figure}
\epsfig{file=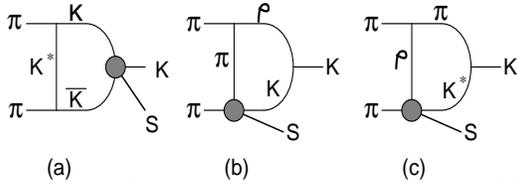,width=7cm,height=2.5cm}
\caption{(a) An inelastic FSI diagram. (b) The ISI diagram of the
nearest left-hand singularity. (c) Another ISI diagram, which produces 
the lowest two-body ISI cut in the OM representation.
\label{fig:2}}
\end{figure}

\noindent
\begin{figure}
\epsfig{file=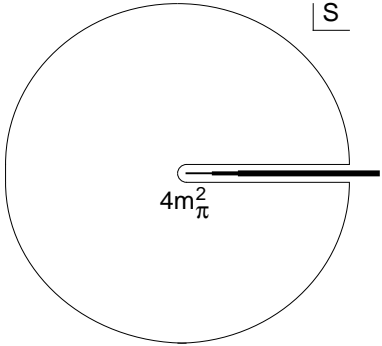,width=5cm,height=4.6cm}
\caption{The $s$-plane singularity of the OM representation.
\label{fig:3}}
\end{figure}

\noindent
\begin{figure}
\epsfig{file=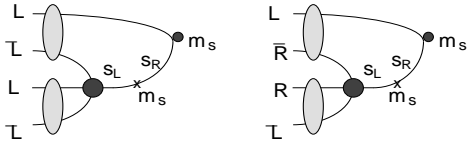,width=7cm,height=2cm}
\caption{The quark diagrams for $K\rightarrow\pi\pi$ through 
(a) the decay operator which contains only left-handed currents and 
(b) the decay operator which contains both left and right currents.
$L(\overline{L})$ and $R(\overline{R})$ denote the left-chiral and
the right-chiral quark, respectively, of $u,d$ flavors  (their 
anti-particles).
\label{fig:4}}
\end{figure}

\noindent
\begin{figure}
\epsfig{file=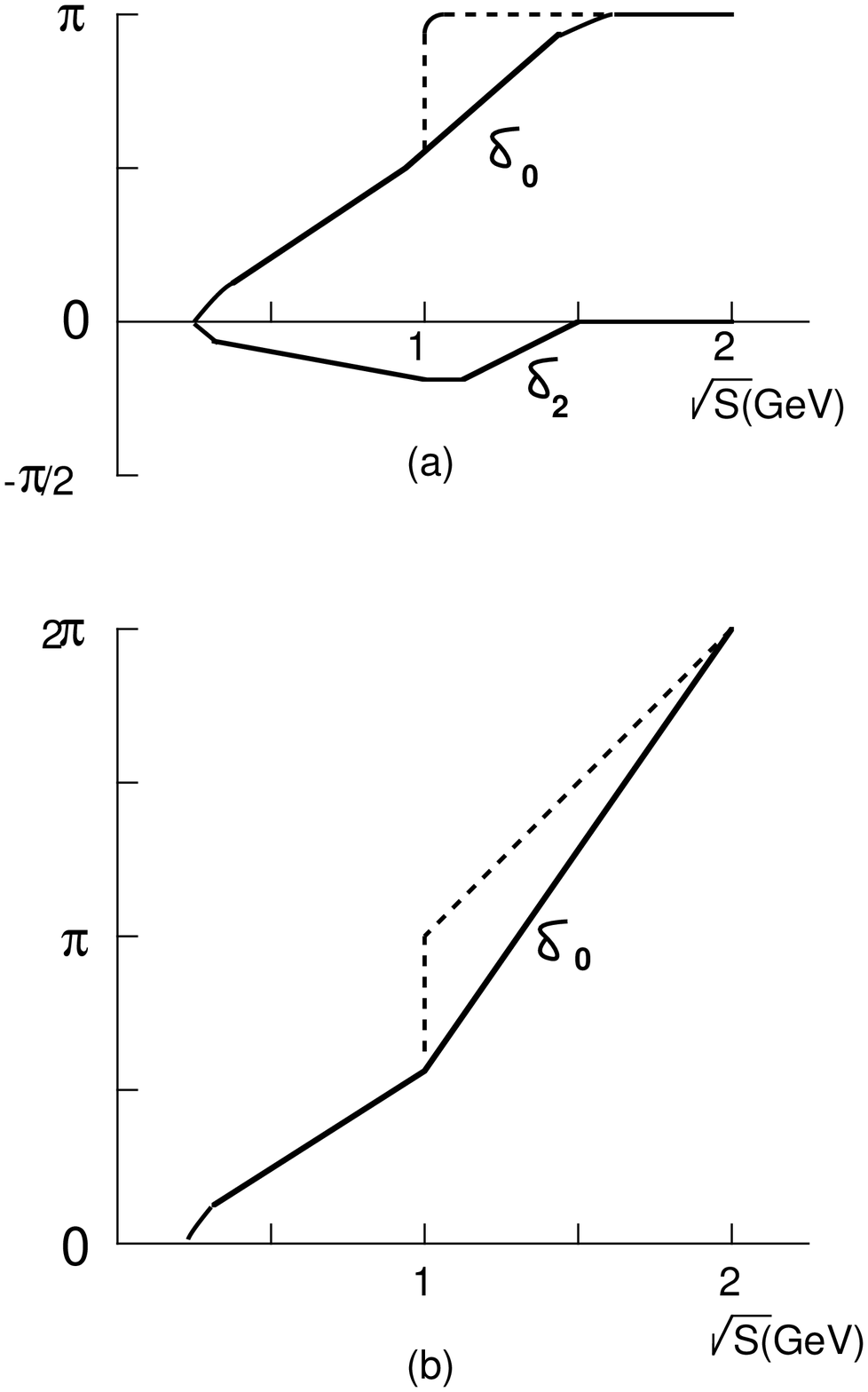,width=7cm,height=10cm}
\caption{(a) A schematic drawing of the behaviors of $\delta_I(s)$ 
for $I=0$ and $I=2$ with the asymptotic values $\delta_0(\infty)=\pi$ 
and $\delta_2(\infty)=0$, respectively. The broken curve is for the
case where $\delta_0(s)$ behaves like the $\pi\pi$ phase shift across
$\sqrt{s} = 2m_K$. (b) $\delta_0(s)$ in the case of $\delta_0(\infty)=
2\pi$ ($N$=2). 
\label{fig:5}}
\end{figure}
 
\end{document}